\documentclass[
    ,final            
  ]
  {aipproc}

\layoutstyle{6x9}
\usepackage{amssymb}

\begin{document}
\vspace*{-25mm}\begin{flushright}
SHEP-09-22  \\  DFTT 62/2009
\end{flushright}
\title{Phenomenological Consequences of the Constrained Exceptional Supersymmetric Standard Model}


\author{Peter Athron}{
  address={Institut f\"ur Kern- und Teilchenphysik, TU Dresden, Dresden, D-01062, Germany}
}

\author{S.~F.~King}{
  address={University of Southampton, Southampton, SO17 1BJ, UK}
}

\author{D.~J.~Miller}{
  address={University of Glasgow, Glasgow, G12 8QQ, UK}
 }

\author{S.~Moretti}{
  address={University of Southampton, Southampton, SO17 1BJ, UK}, altaddress={Dipartimento di Fisica Teorica, Universit\`a di Torino, 
Via Pietro Giuria 1, 10125 Torino, Italy}
}

\author{R. Nevzorov}{
  address={University of Glasgow, Glasgow, G12 8QQ, UK}
}

\begin{abstract}
 The Exceptional Supersymmetric Standard Model (E$_6$SSM) provides a low energy alternative to the MSSM, with an extra gauged U(1)$_N$ symmetry, solving the $\mu$-problem of the MSSM. Inspired by the possible embedding into an E$_6$ GUT, the matter content fills three generations of E$_6$ multiplets, thus predicting exciting exotic matter such as diquarks or leptoquarks. We present predictions from a constrained version of the model (cE$_6$SSM), with a universal scalar mass $m_0$, trilinear mass $A$ and gaugino mass $M_{1/2}$. We reveal a large volume of the cE$_6$SSM parameter space where the correct breakdown of the gauge symmetry is achieved and all experimental constraints satisfied. We predict a hierarchical particle spectrum with heavy scalars and light gauginos, while the new exotic matter can be light or heavy depending on parameters. We present representative cE$_6$SSM scenarios, demonstrating that there could be light exotic particles, like leptoquarks and a U(1)$_N$ Z' boson, with spectacular signals at the LHC.
\end{abstract}

\maketitle


\section{Introduction}

The E$_6$SSM\cite{King:2005jy,King:2005my} is an E$_6$ inspired model with an extra gauged U(1)$_N$ symmetry at low energies, defined by U(1)$_N = 1/4$U(1)$_\chi + \sqrt{15}/4$U(1)$_\psi$, with U(1)$_\chi$ and U(1)$_\psi$ in turn, defined by the breaking, E$_6\rightarrow$SO(10)$\times$U(1)$_\psi$ and SO(10)$\rightarrow$SU(5)$\times$U(1)$_\chi$. The low energy gauge group of the  E$_6$SSM is then \mbox{SU(3)$\times$ SU(2)$\times$U(1)$_Y\times$U(1)$_N$}. 
 
The matter content fills three generations of $27$plet representations of E$_6$, leading to an automatic cancellation of anomalies. Each $27$plet contains one generation of ordinary matter; singlet fields, $S_i$; up and down type Higgs like field, $H_{2,i}$ and $H_{1,i}$ and exotic squarks, $D_i$, $\bar{D}_i$. The model also contains two extra SU(2) doublets, $H'$ and $\bar{H}'$, which are required for gauge coupling unification.  

To evade rapid proton decay we introduce either a $Z_2^B$ or $Z_2^L$ symmetry which work like R-parity except that the exotic quark is odd while it's scalar partners are even.  Under $Z_2^B$ the exotic quarks are leptoquarks while under $Z_2^L$ they are diquarks.  To evade large Flavour Changing Neutral Currents, we also introduce an approximate $Z_2^H$ symmetry, where the Higgs superfields are even, and all others are odd. All couplings through which the exotic quarks and inert Higgs decay violate $Z_2^H$, so the symmetry must only be approximate, but for our Renormalisation Group (RG) analysis they can be neglected. 

Finally so that only the third generation gets VEVs we assume a hierarchical Yukawa sector. Keeping only the dominant couplings the superpotential $W_{E_6SSM} \approx \lambda_iSH_{1,i}H_{2,i}+\kappa_i S D_i\bar{D}_i+h_t H_u Qt^c + h_b H_d Q b^c + h_{\tau} H_d L \tau^c$. $H_u = H_{2,3}$ and $H_d = H_{1,3}$ and $S= S_3$ develop VEVs, $\langle H^0_u \rangle = v_u$,  $\langle H^0_d \rangle = v_d$, giving mass to ordinary matter while $\langle S \rangle = s$ gives exotic quark masses, $\kappa_i S \rightarrow \kappa_i s = \mu_{D_i}$ and an effective $\mu$-term, $\mu_{eff} = \lambda_3 s$. 

\section{The constrained E$_6$SSM}
The constrained E$_6$SSM (cE$_6$SSM) is defined by applying universality constraints to the E$_6$SSM at the scale where the gauge couplings unify, $M_X$. For all scalar ($m_i$), gaugino ($M_i$) and trilinear ($A_i$) masses we have, $ M_i(M_X) = M_{1/2},\;$ $A_i(M_X) = A$ and $m_i(M_{X}) = m_0.$   

 To connect these highscale constraints with low energy phenomenology we employ the RG Equations (RGEs) of the E$_6$SSM given in Ref.~\cite{Athron:2009bs}. Due to the presence of new exotic colored matter the RGE for the strong gauge coupling vanishes at 1-loop, so we use 2-loop RGEs for gauge and Yukawa couplings. We also employ 2-loop RGEs for the gaugino masses and trilinear couplings but only 1-loop RGEs for the soft scalar masses.

 The RGEs for the gauge and Yukawa couplings are independent of the soft breaking masses, but nonlinear even at 1-loop while the soft SUSY breaking sector depends on the gauge and Yukawa couplings as well as the soft SUSY breaking masses but have a simple enough structure that they can be solved semi-analytically to give, \begin{eqnarray}
m_i^2(Q) &=& a_i(Q)  M_{1/2}^2 + b_i(Q) A_0^2 + c_i(Q) A_0 M_{1/2} + d_i(Q) m_0^2,\\
A_i(Q) &=& e_i(Q) A_0 + f_i(Q) M_{1/2},\quad M_i(Q) = p_i(Q) A_0 + q_i(Q) M_{1/2},
\end{eqnarray} where the coefficients depend not only on renormalisation scale, $Q$, but also on the gauge and Yukawa couplings and can be determined numerically for a given $Q$ by selectively setting $M_{1/2}$, $m_0$ and $A$ to zero and evolving between $M_X$ and $Q$ with the full set E$_6$SSM RGEs. Unlike the constrained MSSM, in the cE$_6$SSM we find that in contrast with the cMSSM RG coefficients obey $p_j, q_j \lesssim  a_i, d_i$, for all $j$ and all $i$ from ordinary matter.  This implies a gaugino sector which is light in comparison to the sfermions.

To make physical predictions we then combine these with the electroweak symmetry breaking (EWSB) conditions $\partial V/ \partial s = \partial V/ \partial v_1 =  \partial V/ \partial v_2 = 0$, where $V$ is the Higgs potential including leading 1-loop stop contributions. This gives $M_{1/2}$, $A$ and $m_0$ values consistent with both the highscale universality conditions and the EWSB.  Finally we calculate the physical masses and test against experimental constraints.  We require a mass bound of $300$ GeV for the squarks, gluinos, exotic quarks and squarks; $100$ GeV for the inert Higgs and Higgsinos and 860 GeV for $Z^\prime$ boson. We also keep Yukawa couplings less than $3$ to maintain perturbativity and insist on a neutralino LSP.    
\section{Results}
\begin{figure}[h!]
\begin{tabular}{cc}
\resizebox{!}{5cm}{%
\includegraphics{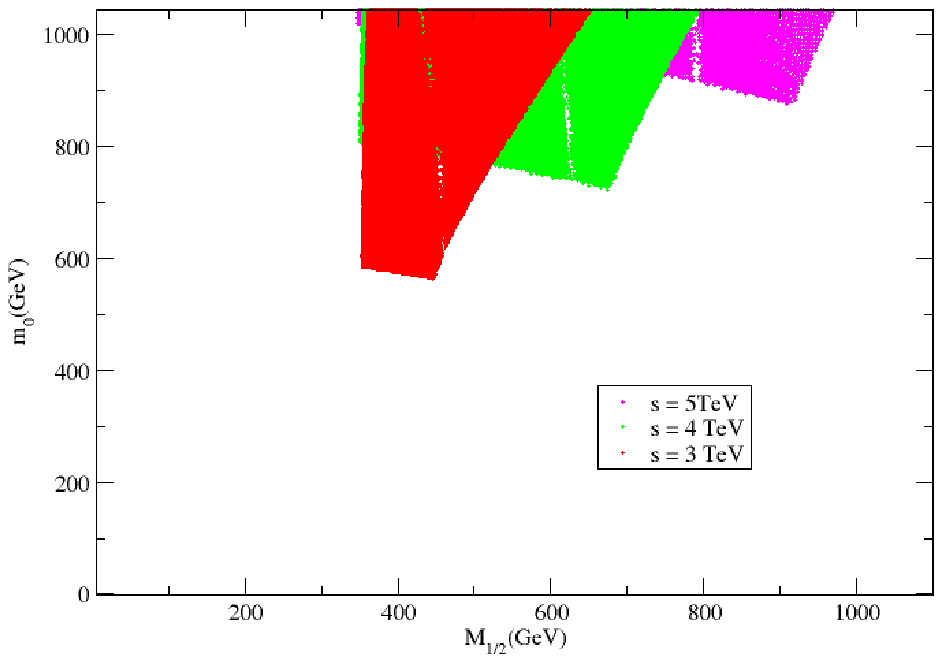}} &
\resizebox{!}{5cm}{\includegraphics{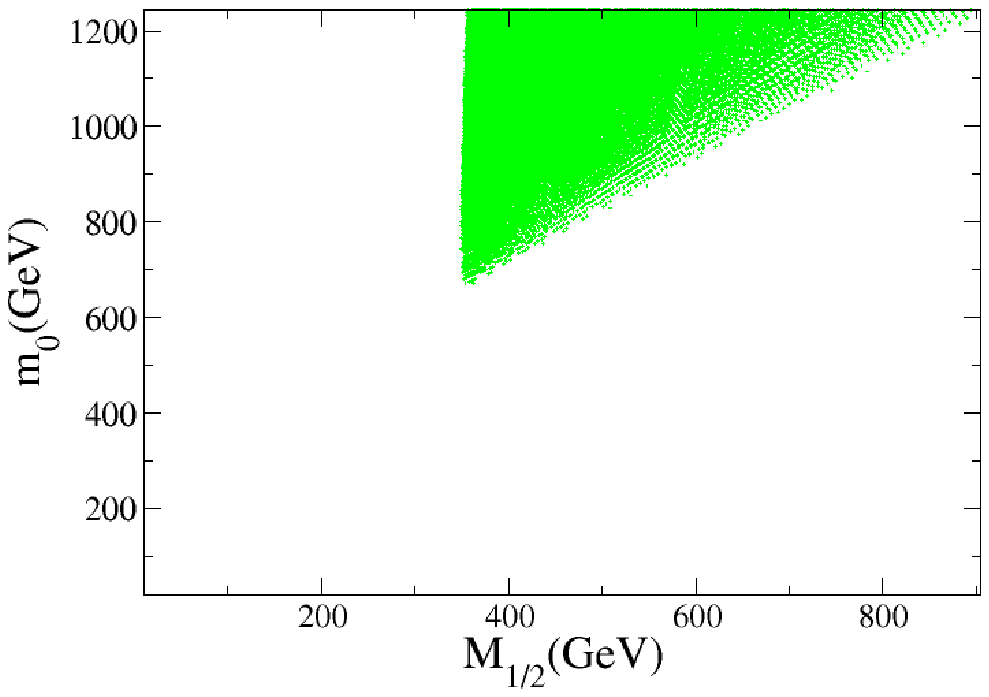}}
\caption{Physical solutions with $\tan \beta = 10$, $\lambda_{1,2} = 0.1$, $s= 3, 4, 5$ TeV fixed and $\lambda \equiv \lambda_3$ and
$\kappa \equiv \kappa_{1,2,3}$ varying (left), and letting $s$ also
vary (right), which pass experimental constraints from LEP and
Tevatron data.\label{tb10_Valid}}
\end{tabular}
\end{figure}

As shown in Fig.~\ref{tb10_Valid}(left) for fixed values of $s$ we find many phenomenologically acceptable solutions at the TeV scale by varying the Yukawa couplings $\lambda$ and $\kappa$.  Notice also that although $m_0$ and $M_{1/2}$ vary with the Yukawas, in general the mass scale increases with singlet VEV $s$.  Scanning over $s$ too, Fig.~(\ref{tb10_Valid},right) we see that $m_0 \gtrsim M_{1/2}$, which further pushes up the masses of the sfermions and in combination with the comparative magnitudes of the RG coefficients implies that all the sfermions of ordinary matter are heavier than the gluino, the lightest two neutralinos and the lightest chargino. 

\begin{figure}[h]
\begin{tabular}{cc}
\resizebox{!}{6cm}
{\includegraphics{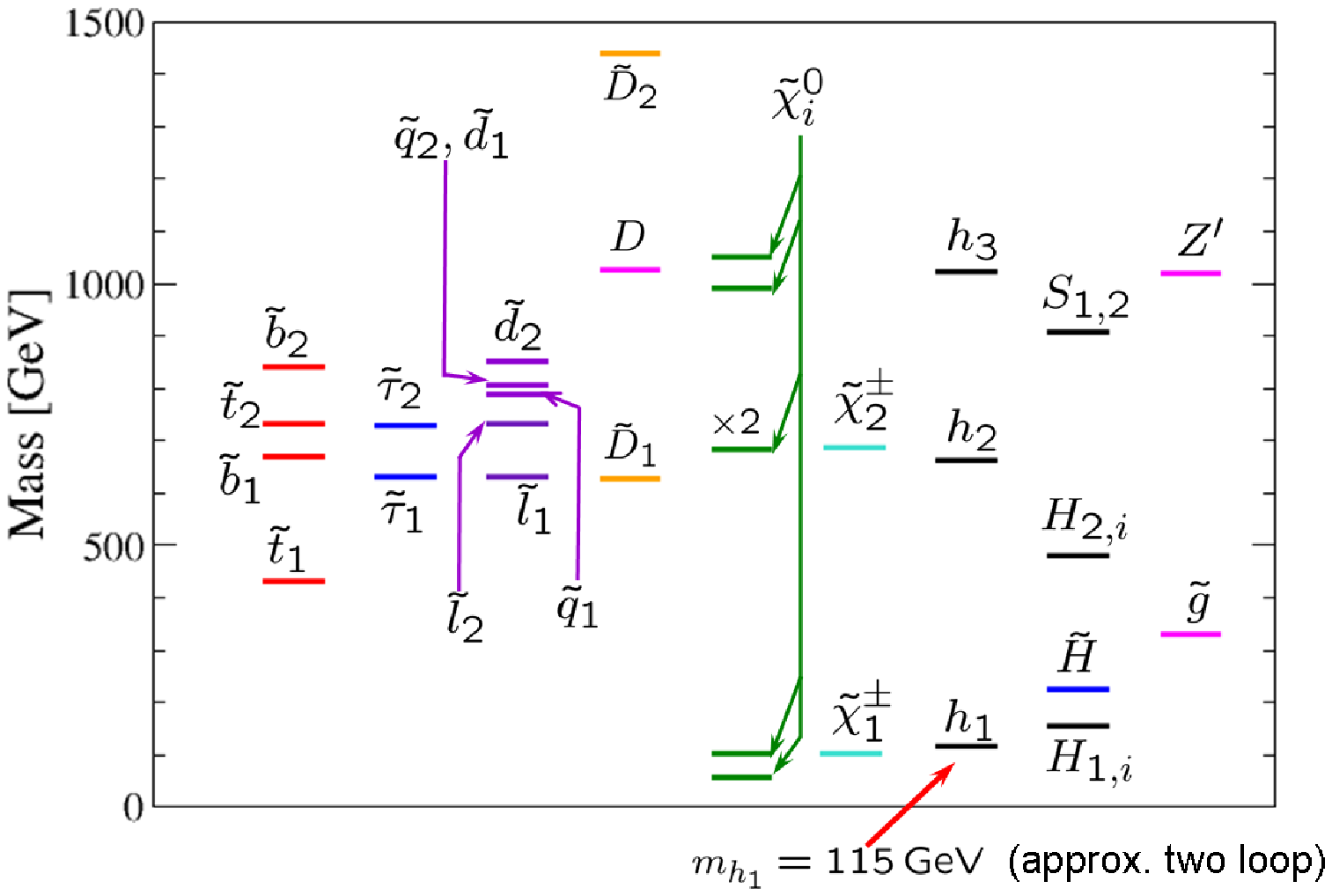}} &
\resizebox{!}{6cm}
{\includegraphics{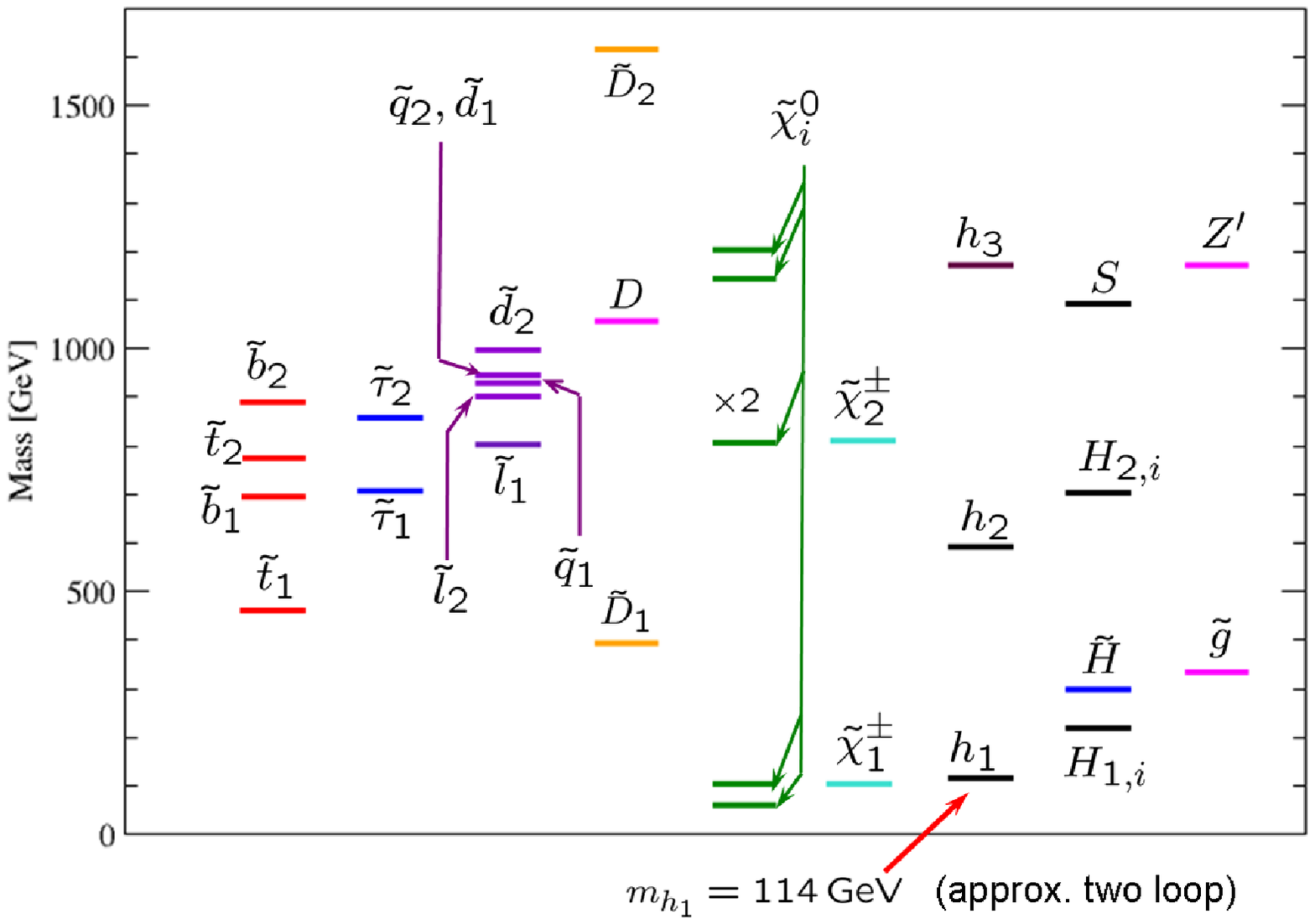}} \\
\resizebox{!}{6cm}
{\includegraphics{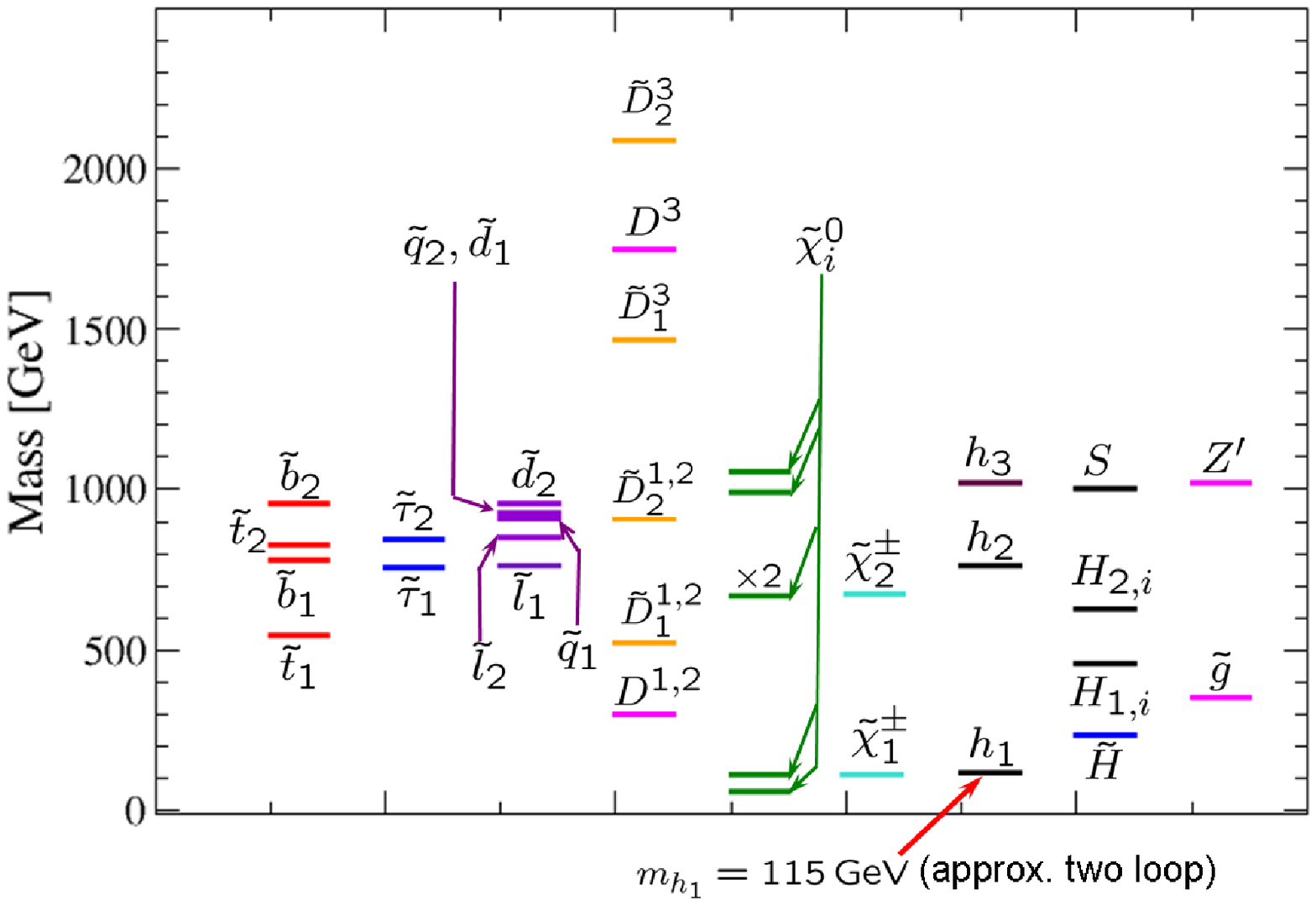}}&
\resizebox{!}{6cm}{%
\includegraphics{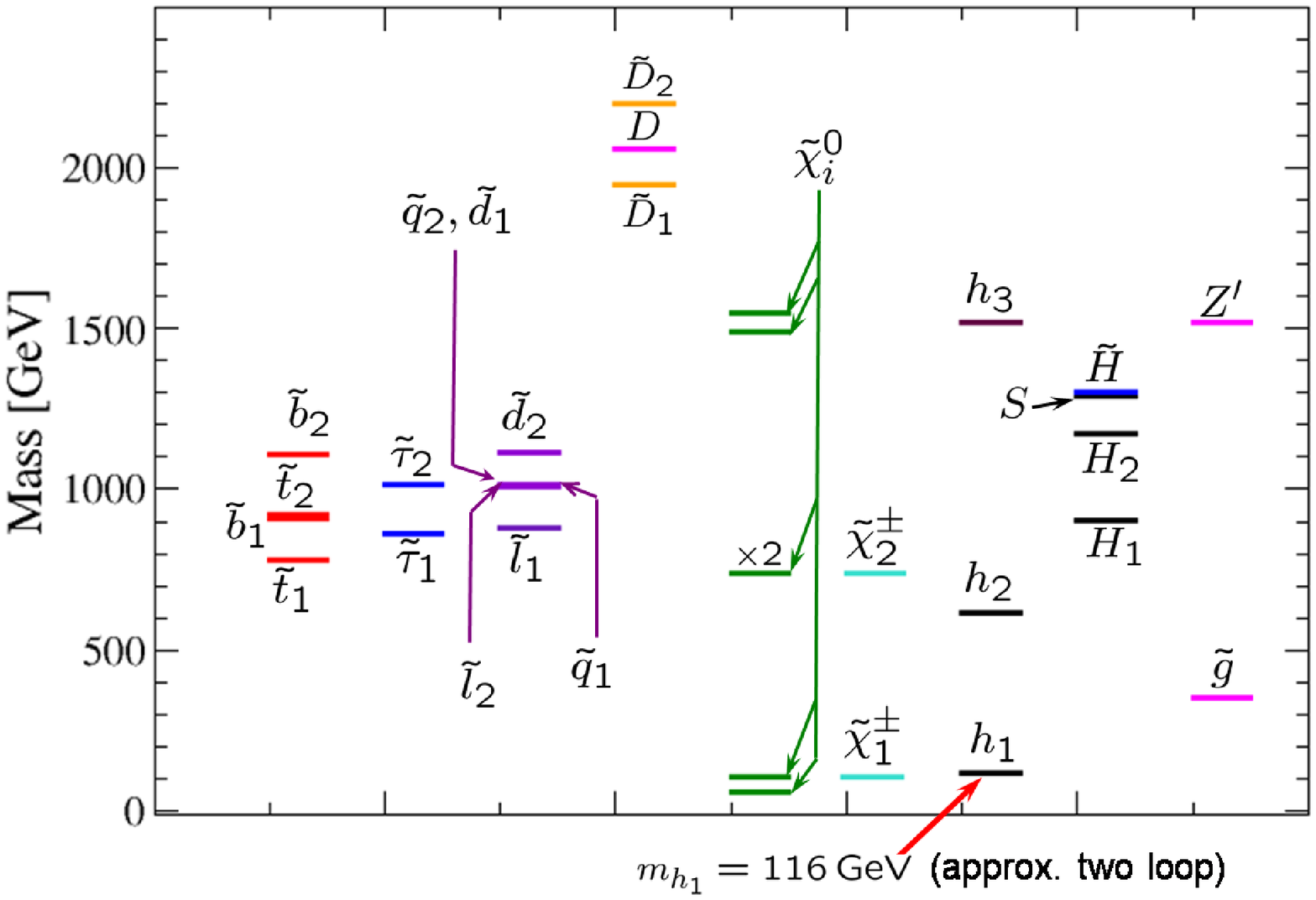}}\\
\end{tabular}
\caption{Mass spectra for scenarios 1 (top left), 2 (top right), 3 (bottom left), 4 (bottom right). \label{benchmarks} }
\end{figure}

We also present representative scenarios in Fig.~\ref{benchmarks},
showing the range of possible signatures that could be seen at the
LHC. Scenario 1, top left, is drawn from the bottom left
corner of Fig.~\ref{tb10_Valid}(right) so that $m_0$ and $M_{1/2}$ are
as light as possible. Nonetheless all the sfermions of ordinary matter
are rather heavy, with only the lightest stop below $500$ GeV.  As
well as the characteristic light gaugino sector and the light Higgs,
we also have light Inert Higgs bosons and the Inert Higgsinos.  The Inert Higgs bosons decay via $Z_2^H$ violating terms that
are analogous to the Yukawa interactions of the Higgs superfields,
${H}_u$ and $H_d$.  So the inert Higgs bosons decay predominantly into
3rd generation fermion--anti-fermion pairs like $H^{0}_{1,\,i}
\rightarrow b \bar b$ for neutral states  or $H^{-}_{1,\,i} \rightarrow \tau
\bar{\nu}_{\tau}$ for charged states.   Similarly the inert Higgsinos decay into fermion-anti-sfermion pairs, e.~g.~$\tilde H_{i}^0 \rightarrow t
\tilde{\bar{t}}^*$. 
 
In scenario 1 all exotics squarks are heavy due to the large singlet
VEV contribution to their mass, but mixing effects can render one of
these masses light as in scenario 2, top right. With a universal
$\kappa$ coupling though all exotic quarks must be heavy as a large
$\kappa_i$ is required to drive EWSB.  However if we split the
$\kappa_i$ couplings then only one generation need be heavy, as in scenario 3, bottom left, giving rise to a remarkable signature.

Exotic quarks also decay through $Z_2^H$ violating couplings and
assuming the third generation couplings dominate, the lightest exotic
quarks decay into states like $\tilde{t}b$, (if diquarks) or
$\tilde{t}\tau$, (if leptoquarks), substantially enhancing the cross
section of either $pp\to t\bar{t}b\bar{b}+E^{\rm miss}_{T}+X$ (if
diquarks) or $pp\to t\bar{t}\tau \bar{\tau}+E^{\rm miss}_{T}+X$ or
$pp\to b\bar{b}+ E^{\rm miss}_{T}+X$ (if leptoquarks). SM production
of $ t \bar t
\tau^+ \tau ^-$ is $(\alpha_W / \pi)^2$ suppressed in comparison to
the leptoquark decays, so light leptoquarks should produce a strong
signal at the LHC. Similarly scalar leptoquarks
decay into quark--lepton final states like $\tilde D
\rightarrow t \tau$, and pair production leads to an enhancement of
$pp \rightarrow t \bar t \tau \bar{\tau}$ (without missing energy) at
the LHC.

 Although such scenarios are phenomenologically exciting it is not
 guaranteed that any new matter from the 27plets is very light.  All
 such particles could be rather heavy and challenging to detect, like
 scenario 4, bottom right, but even in such a pessimistic case there
 is still the striking prediction of the light gluino. Due to the
 hierarchical spectrum, the gluinos can be relatively narrow states
 with width $\Gamma_{\tilde{g}}\propto
 M_{\tilde{g}}^5/m_{\tilde{q}}^4$, comparable to that of $W^{\pm}$ and
 $Z$ bosons.  They will decay via $\tilde{g} \rightarrow q
\tilde{q}^* \rightarrow q \bar{q} + E_T^{\rm miss}$, so gluino pair
production implies an appreciable enhancement of the cross
section for $pp \rightarrow q \bar q q \bar q + E_T^{\rm miss} + X$,
where $X$ refers to any number of light quark/gluon jets.


\begin{theacknowledgments}
RN acknowledges support from the SHEFC grant HR03020 SUPA 36878.
SM is financially supported in part by the scheme `Visiting Professor - Azione D - 
Atto Integrativo tra la Regione Piemonte e gli Atenei Piemontesi.
\end{theacknowledgments}



  



\bibliographystyle{aipproc}   


\bibliography{sample}

\IfFileExists{\jobname.bbl}{}
 {\typeout{}
  \typeout{******************************************}
  \typeout{** Please run "bibtex \jobname" to optain}
  \typeout{** the bibliography and then re-run LaTeX}
  \typeout{** twice to fix the references!}
  \typeout{******************************************}
  \typeout{}
 }

\end{document}